# Side-wall pressure distribution of granular silos

Yusheng Lei[1,2], Qinwei Ma[2,*], Qingfan Shi[1,*]

1) School of Physics, Beijing Institute of Technology, Beijing 100081

2) School of Aerospace Engineering, Beijing Institute of Technology, Beijing 100081

*These authors should be considered as co-corresponding authors. E-mail: maqw@bit.edu.cn and

qfshi123@bit.edu.cn

**ABSTRACT**

The side-wall pressure distribution in static silos is an important and open problem related to storage safety of granular materials. Referring to Lvin's theory [Powder Technology 4, 280 (1971)] and Rodolfo's 2D numerical result [Phys. Rev. E 97.1(2018)], a new theoretical model of side-wall pressure is established by assuming the friction coefficient as well as the ratio of horizontal to vertical stress changing with depth position in silo. Furthermore, the transition of stress state between hydrostatic-like and Janssen-like is systematically investigated by our model and experiment, and finally an empirical expression describing this transition is proposed.

## 1. INTRODUCTION

Silos are frequently used to store granular material for the food industry, agriculture, pharmacy and chemistry, yet some silos collapse each year [1, 2]. These serious accidents that result in a great number of economic loss stem from the design defects of silos. Failure in design points to the fundamental difficulties in accounting correctly for the structural properties of silos and the physics of granular materials altogether. So, the understanding of pressure on the bottom and lateral wall is critical to the design of silos. In fact, the distribution of stress in silos is still a problem worthy of further research until today though many pioneers studied this problem since nineteenth century.

Huber-Burnand initiated the studies of force repartition in granular materials through an experiment in which an egg covered with a thick layer of sand was able to support a heavy block of iron without breaking [3]. Later, Rankine established an early representative theory that the pressure in silos varies linearly with depth position [4]. In 1895, based on the facts that stress propagation in granular column are seldom isotropic and the friction play an important role in jammed granular materials, Janssen developed a simple model to explain how stresses are acted in a silo [5]. Especially, this model has two basic assumptions: the ratio of horizontal to vertical stress is the same value at any point; the effect of silo bottom is neglected, i.e. silo is assumed to be infinitely deep. Though Janssen's model can precisely give a description of the apparent mass changing exponentially with depth, it is failed to explain many experiments such as the apparent mass of a silo with extra load [6-8]. Similarly, can Janssen theory correctly reflect the side-wall pressure of a silo? In fact, Janssen's model is no longer valid for the side-wall pressure of filling and discharging state of silo [9-11]. Furthermore, whether Janssen's model is suitable for the case that granular column has a quasi-static descent displacement after initial filling? A most recent simulation performed in a poly-dispersed two-dimensional column after such a "descent displacement" by using molecular dynamics validate the existence of the point of maximum stress and find a hydrostatic-like phenomenon of stress [12]. Note that the operation of a descent displacement after initial filling of silo is important in physics, since it could make the packing fraction tending to a stable value in the silo [6, 13]. Even if in the perspective of engineering, the state of the silo after such descent displacement is similar to the silo after a partial discharging or a sudden disturbing. So, the peak value and the hydrostatic-like phenomenon of side-wall pressure appeared in this kind static silo are worthy of further investigation theoretically and experimentally.

In this paper, we study side-wall pressure distribution based on Lvin's theory of static silos and modification of Janssen hypothesis [12, 14]. Considering the hydrostatic-like behavior in upper region and the geometric restriction in bottom region, we first assume the friction coefficient and the ratio of horizontal to vertical stress changing with depth position, and then substitute it into the differential equation of inner stress, finally deduce a non-Janssen expression of side-wall pressure in which a peak exists. In addition, given that the stress-state transition between hydrostatic-like to Janssen-like is related to the geometric parameters of the granular column, we investigate how these parameters in our model impact on the distribution of side-wall pressure and transferring ratio in our model. After that, we combine the experimental observations and the theoretical analysis of the state transition existed in vertical stress distribution, an empirical expression describing the stress state transition is obtained.

## 2. Theoretical analysis

Above all, a brief review of Janssen's theory is necessary. A differential equation by using the ultimate equilibrium condition to a uniformly loaded elementary disc is derived as the follows

$$\frac{dp_v}{dy} + \frac{2k}{R} p_v = \gamma, \tag{1}$$

where $p_v$ is the vertical stress, $y$ is the depth position, $\gamma$ is the specific weigh, $R$ is the radius of silo, and $k$ is generally defined as $fJ_0$. Furthermore, $f$ means the internal friction coefficient, $J_0$ represents Janssen's stress ratio (i.e. the ratio of horizontal to vertical stress) and is expressed by

$$J_0 = \frac{1 - sin\phi}{1 + sin\phi}, \tag{2}$$

where $\phi$ is the internal friction angle of the glass beads. By solving eq. (1) with the boundary condition $p_v|_{y=0} = 0$, we can get the horizontal stress on side wall $p_h$ (i.e. wall pressure) as

$$p_h = J_0 p_v = \frac{\gamma R}{2f}\left(1 - exp\left(-\frac{2k}{R}y\right)\right), \qquad (3)$$

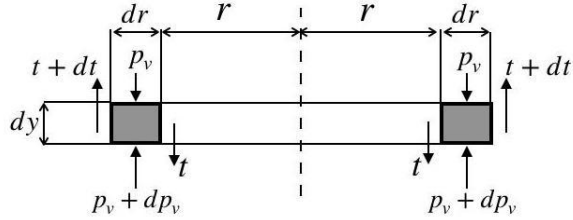

Fig. 1 Schematic of stress analysis in Lvin's model

Different from Janssen's method, Lvin applied the ultimate equilibrium condition to the elementary ring and assumed that all the inside rings slide downward relatively to the outside ones (Fig 1). In this way a partial differential equation was derived as follows

$$\frac{\partial P_v}{\partial y} + k\left(\frac{\partial p_v}{\partial r} + \frac{p_v}{r}\right) = \gamma, \qquad (4)$$

where $r$ is the inner radius of ring. For the convenience of solving eq. (4), several variables were defined as

$$\rho = r/R;\ \eta = ky/R;\ \psi = kp_v/\gamma R;\ \Psi = \rho\psi. \qquad (5)$$

Then, the eq. (4) can be rewritten as

$$\frac{\partial \Psi}{\partial \eta} + \frac{\partial \Psi}{\partial \rho} = \rho, \qquad (6)$$

where $\eta$ means the dimensionless depth position, $\rho$ is the dimensionless inner radius of ring.

The general solution of eq. (6) is given by

$$\Psi = \frac{1}{2}\rho^2 + \Phi(\eta - \rho) \tag{7}$$

Hence, a family of parallel characteristic lines '$\eta - \rho = C_i$' is introduced based on eq. (7), the arbitrary function $\Phi(\eta - \rho)$ becomes constant on each of these lines. The characteristic line which pass through the origin of coordinates divides the granular column into two regions: the upper one A ($\eta \leq \rho$) including the free surface points, and the lower one B ($\eta \geq \rho$) including the silo axis points. Considering the stresses are zero on the free surface and finite on the axis, the boundary conditions $\Psi|_{\eta=0} = 0 \,\&\, \Psi|_{\eta=0} < \infty$ are determined. Thus, the continuous solution of eq. (4) is

$$\Psi = \begin{cases} \eta\left(\rho - \dfrac{\eta}{2}\right) & \eta \leq \rho \\ \dfrac{1}{2}\rho^2 & \eta \geq \rho \end{cases}. \tag{8}$$

The wall pressure ($\rho = 1$) can be derived as

$$p_h|_{\rho=1} = \frac{\gamma R J_0}{k}\psi = \frac{\gamma R J_0}{k}\Psi|_{\rho=1} = \begin{cases} \dfrac{\gamma R}{f}\eta\left(1 - \dfrac{\eta}{2}\right) & \eta \leq 1 \\ \dfrac{\gamma R}{2f} & \eta \geq 1 \end{cases}, \tag{9}$$

Note that the stress ratio is a constant in the above discussion. In fact, Lvin also considered the condition in which the stress ratio changes with the depth position in some specific cases. For example, in this method the stress distribution of the material between two characteristic horizontal silo sections was analyzed with a non-constant stress ratio assumed, and an analytical formula of side-wall pressure in this transition region was obtained.

Inspired by Lvin's method, we suggest that the propagating pattern of stress at the bottom region of granular column is similar to that of the interface region of two granular columns with different stress ratios. In other words, the stress ratio $J$ is assumed to increase with the depth position in the bottom region of granular column. For intuitive speaking, this assumption can be understood as the transition of stress field in the granular column from the Rankine's active to the passive state. Thus, we can suppose

$$J = \begin{cases} qJ_0 & \eta \leq \eta_0 - \eta_1 \\ qJ_0\left[1 + (m-1)\left(\dfrac{\eta - (\eta_0 - \eta_1)}{\eta_1}\right)^s\right] & \eta_0 \geq \eta \geq \eta_0 - \eta_1 \end{cases}, \quad (10)$$

where, the maximum value of $J$ in the column is $qmJ_0$ ($m>1$) as shown by the red line in Fig. 2(a), $q$, $m$ and $s$ are undetermined parameters related to the stress state, $\eta_1$ is the length of bottom region, and the dimensionless total depth of granular column is $\eta_0 = kH/R$.

Besides, considering the effect of both bottom and side boundary condition (i.e. the Janssen-like tendency) and the hydrostatic-like tendency of stress [10], the friction coefficient $\mu$ can be assumed to increase with the depth position in the upper region of granular column

$$\mu = \begin{cases} f\left(\dfrac{\eta}{\eta_0 - \eta_1}\right)^b & \eta \leq \eta_0 - \eta_1 \\ f & \eta_0 \geq \eta \geq \eta_0 - \eta_1 \end{cases}, \quad (11)$$

where $b$ is also an undetermined parameter which will be discussed later. This assumption can be understood as an average effect at the radial direction caused by a heterogeneous distribution of friction state (see the blue line in Fig. 2(a)). In addition, a crucial coupling factor $K(\eta) = J(\eta)\mu(\eta)$, indicating the stress transferring from the vertical stress to the friction on the wall, is defined and depicted in Fig. 2(b).

Then, we can derive a piecewise equation:

$$\begin{cases} \dfrac{\partial \Psi}{\partial \eta} + q\left(\dfrac{\eta}{\eta_0 - \eta_1}\right)^b \dfrac{\partial \Psi}{\partial \rho} = \rho & \eta \leq \eta_0 - \eta_1 \\ \dfrac{\partial \Psi}{\partial \eta} + q\left[1 + (m-1)\left(\dfrac{\eta - (\eta_0 - \eta_1)}{\eta_1}\right)^s\right]\dfrac{\partial \Psi}{\partial \rho} = \rho & \eta_0 \geq \eta \geq \eta_0 - \eta_1 \end{cases}. \quad (12)$$

For the upper region of $0 \leq \eta \leq \eta_0 - \eta_1$, similar to the procedure that presented above, the general solution is

$$\Psi = \rho\eta - \frac{q\eta^2}{(b+2)}\left(\frac{\eta}{\eta_0 - \eta_1}\right)^b + \Gamma\left(\rho - \frac{q\eta}{(b+1)}\left(\frac{\eta}{\eta_0 - \eta_1}\right)^b\right). \tag{13}$$

Next we consider the boundary conditions on the surface and central axis: $\Psi|_{\eta=0} = 0$ & $\Psi|_{\eta=0} < \infty$, the dimensionless stress in the upper region is obtained:

$$\Psi = \rho\eta - \frac{q\eta^2}{(b+2)}\left(\frac{\eta}{\eta_0 - \eta_1}\right)^b. \tag{14}$$

For the bottom region of $\eta_0 - \eta_1 \leq \eta \leq \eta_0$, the general solution is

$$\Psi = \left[\rho(\eta - \eta_0 + \eta_1) - \frac{q(\eta - \eta_0 + \eta_1)^2}{2} - q(m-1)\frac{(\eta - \eta_0 + \eta_1)^{s+2}}{(s+2)\eta_1^s}\right] + \\ \Theta\left(\rho - q(\eta - \eta_0 + \eta_1) - q(m-1)\frac{(\eta - \eta_0 + \eta_1)^{s+1}}{(s+1)\eta_1^s}\right), \tag{15}$$

where $\Theta$ is the arbitrary function. Accordingly, we can first get the family of characteristic lines determined by the eq. (15) as the follows

$$\rho - q(\eta - \eta_0 + \eta_1) - q(m-1)\frac{(\eta - \eta_0 + \eta_1)^{s+1}}{(s+1)\eta_1^s} = C_i. \tag{16}$$

Taking Lvin's analysis method for the length of "the region between the stabilization level and rupture level", we can calculate the length of bottom region $\eta_1$ by analyzing eq. (16). Here, assume the characteristic line passing through the bottom corner of silo, then consider the cross point of the characteristic line. Thus, the length of bottom region can be determined by the height of the cross point as shown in Fig. 2(c). Finally, we get

$$\eta_1 = \frac{s+1}{q(m+s)}. \tag{17}$$

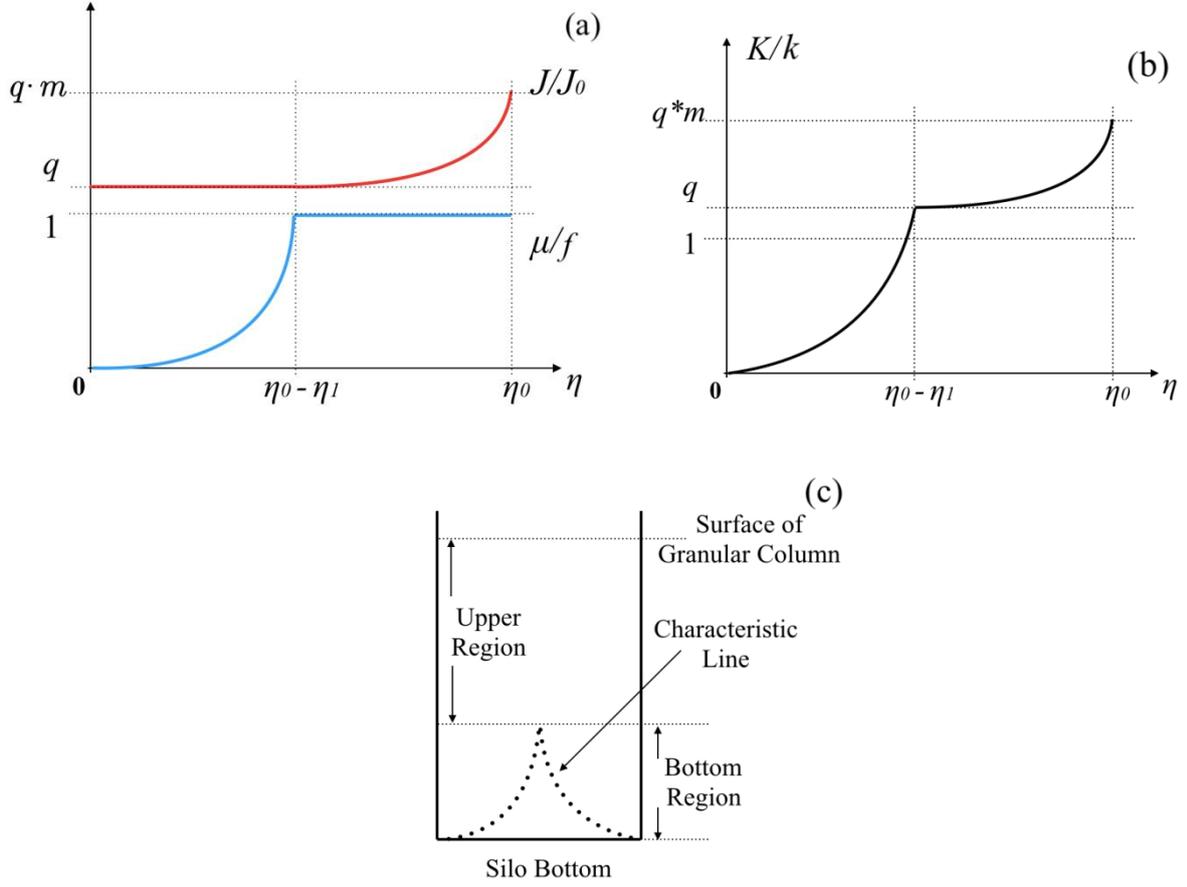

Fig. 2 Schematic of (a) stress ratio $J/J_0$ and stress ratio $\mu/f$, (b) transferring ratio $K/k$, (c) bottom region

As for the boundary condition of eq. (11), considering the engineering practice, we have $\Psi|_{\eta=\eta_0-\eta_1} = \rho(\eta_0-\eta_1) - q(\eta_0-\eta_1)^2/(b+2)$ and $\Psi|_{\rho=0} < \infty$. So, we can derive the continuous solution of eq. (11) as the follows

$$\Psi = \begin{cases} \eta\rho - \dfrac{q\eta^2}{(b+2)}\left(\dfrac{\eta}{\eta_0-\eta_1}\right)^b & \eta \leq \eta_0 - \eta_1 \\ T_1 + T_2 & \eta_0 \geq \eta \geq \eta_0 - \eta_1 \end{cases}, \qquad (18)$$

where

$$T_1 = \left[\rho(\eta-\eta_0+\eta_1) - \frac{q}{2}(\eta-\eta_0+\eta_1)^2 - q\left(\frac{m-1}{s+2}\right)\frac{(\eta-\eta_0+\eta_1)^{s+2}}{\eta_1^s}\right],$$

$$T_2 = \left[(\eta_0 - \eta_1)\left(\rho - q(\eta - \eta_0 + \eta_1) - q\left(\frac{m-1}{s+1}\right)\frac{(\eta - \eta_0 + \eta_1)^{s+1}}{\eta_1^s} - \frac{q(\eta_0 - \eta_1)}{(b+2)}\right)\right].$$

So the wall pressure ($\rho = 1$) can be derived as

$$p_h\big|_{\rho=1} = \frac{\gamma R J}{k}\Psi\big|_{\rho=1} = \begin{cases} \dfrac{q\gamma R}{f}\left[\eta - \dfrac{q\eta^2}{b+2}\left(\dfrac{\eta}{\eta_0 - \eta_1}\right)^b\right] & \eta \leq \eta_0 - \eta_1 \\ \dfrac{jq\gamma R}{f}\left(T_1\big|_{\rho=1} + T_2\big|_{\rho=1}\right) & \eta_0 \geq \eta \geq \eta_0 - \eta_1 \end{cases}, \qquad (19)$$

where

$$j = \left[1 + (m-1)\left(\frac{\eta - \eta_0 + \eta_1}{\eta_1}\right)^s\right].$$

Since the parameters $q$, $s$ and $m$ in our model may be various for different granular columns, such as it may correlate with the diameter ratio of granular particles with silo. For the sake of intuition, based on eq. (19) we picture out the effect of changing $q$, $b$ and $m$ (parameter $s$ is omitted because it just served as a smooth factor in the model) on the wall pressure and the transferring factor $K(\eta)$ as shown in figs. 3 and 4.

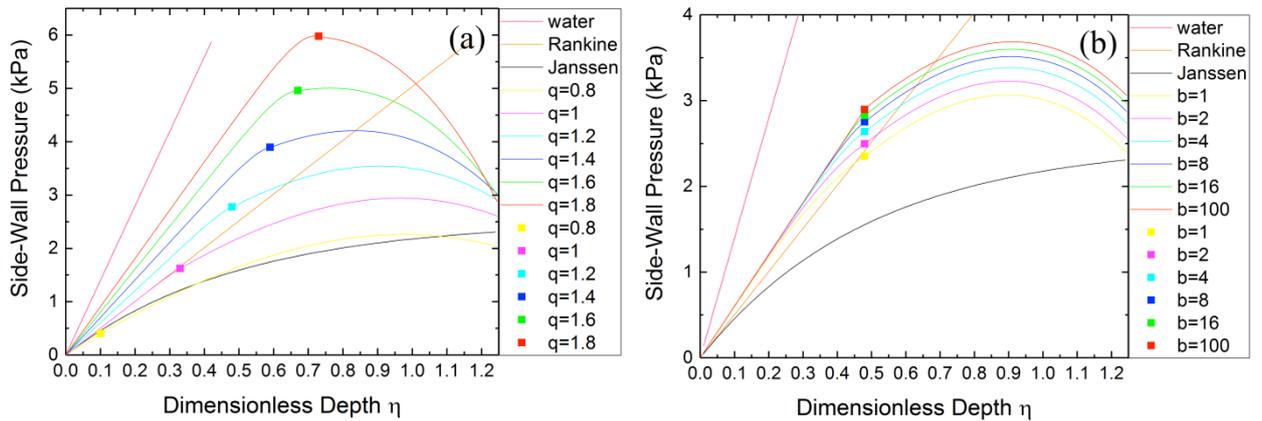

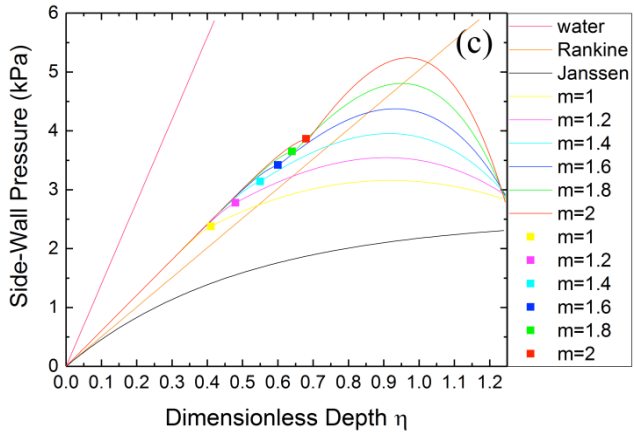

Fig. 3  Side-wall pressure: (a) Change $q$ from 0.5 to 1.5 when $s=1$, $m=1.2$, $b=10$, $D=190$mm, $H=730$mm; (b) Change $b$ from 1 to 16 when $s=1$, $m=1.2$, $q=1.2$, $D=190$mm, $H=730$mm; (c) Change $m$ from 1 to 2 when $s=1$, $b=10$, $q=1.2$, $D=190$mm, $H=730$mm. (For the hydrostatic pressure curve: $p_h=\gamma\eta R/k$; and for Rankine's theoretical curve [4] ).

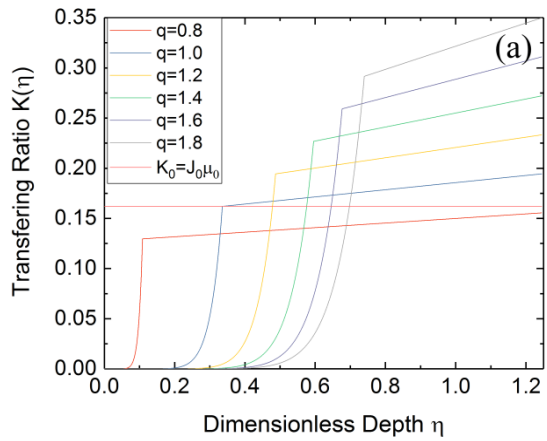
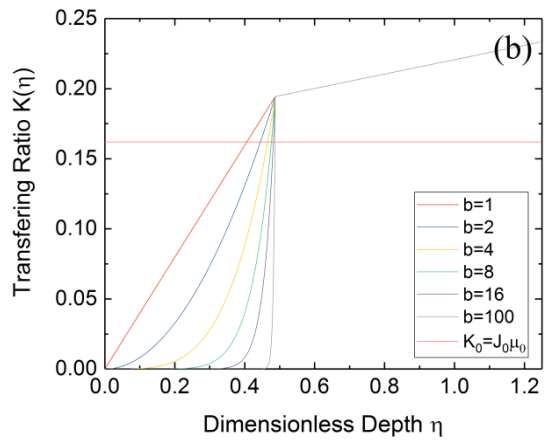
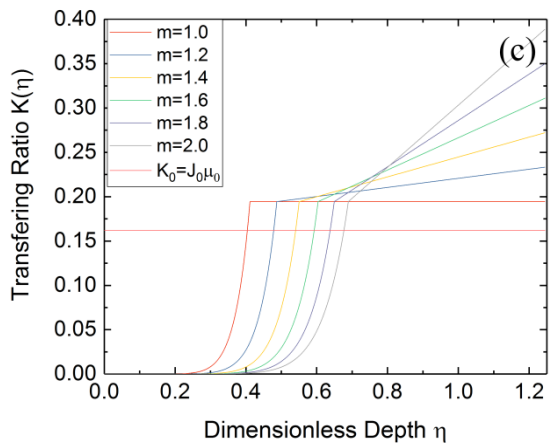

Fig. 4 Transferring ratio $K$: (a) Change $q$ from 0.5 to 1.5 when $s=1$, $m=1.2$, $b=10$, $D=190$mm, $H=730$mm; (b) Change $b$ from 1 to 16 when $s=1$, $m=1.2$, $q=1.2$, $D=190$mm, $H=730$mm; (c) Change $m$ from 1 to 2 when $s=1$, $b=10$, $q=1.2$, $D=190$mm, $H=730$mm.

Obviously, the bigger $q$ is, the higher peak location of wall pressure distribution as well as the greater closeness of wall pressure curve to the hydrostatic stress is (see Fig. 3 (a)). Indeed, the increment of $q$ results in the increasing of transferring ratio $K$ in the upper region of granular column (see Fig. 4 (a)) as well as decreasing of length of bottom region as shown by square points in Fig. 3 (a).

Besides, the bigger $b$ is, the higher peak location of wall pressure is (see Fig. 3 (b)). However, the effect of $b$ is certainly not that noticeable as $q$, since $b$ is just a parameter telling how fast the friction coefficient approaching to $\mu_0$ (see Fig. 4(b)). Furthermore, $b$ do not influence much on the overall approaching tendency of how the wall pressure approach to the hydrostatic state.

From Fig. 3 (c), the bigger $m$ is, the larger peak value of wall pressure distribution is. This result is quite different from the effect caused by $q$, because, the changing of $m$ mainly affects the stress state (see Fig. 4 (c)) of bottom region rather than the hydrostatic-like state occurred in upper region.

Now, combining the analysis above and geometric parameters in our model, a new factor $\lambda$ is induced to weight the level of the hydrostatic-like effect and the transition of stress state:

$$\lambda = \frac{aH}{R}, \qquad (20)$$

where $a$ is the radius of granular beads.

Note that the factor $\lambda$ is closely related to the length ratio of bottom region vs. total granular column (i.e. $\eta_1/\eta_0$). In order to understand their relationship, we take two extreme situations here as examples: firstly, when the upper region is almost disappeared ($\eta_1/\eta_0 \to 1$) the non-hydrostatic-like stress state is in dominant, which is equivalent to $\lambda \to 0$, i.e., $a/R$ is not too small and $H$ is large enough. Secondly, when the upper region nearly covers all silo height $\eta_1/\eta_0 \to 0$, the stress in silo is hydrostatic-like state, i.e., when the silo is giant enough meanwhile the bead is relatively small enough (i.e. $\lambda \to 0$), the silo almost presents a hydrostatic pressure distribution. Notice that these conclusions are obtained with totally neglecting the effect of cohesion and friction between beads, if not, then the $\eta_1/\eta_0 \to const. << 1$ when $\lambda \to 0$ and $\eta_1/\eta_0 \to const. >> 0$ when $\lambda \to \infty$. The rule of how $\eta_1/\eta_0$ changes with factor $\lambda$ is depicted in Fig. 5. Furthermore, an empirical expression can be suggested based on this analysis:

$$\frac{\eta_1}{\eta_0} = B\{1 + exp[-A(\lambda + \lambda_0)]\}^{-1}, \tag{21}$$

where $A$, $B$, $\lambda_0$ will be determined by experimental data.

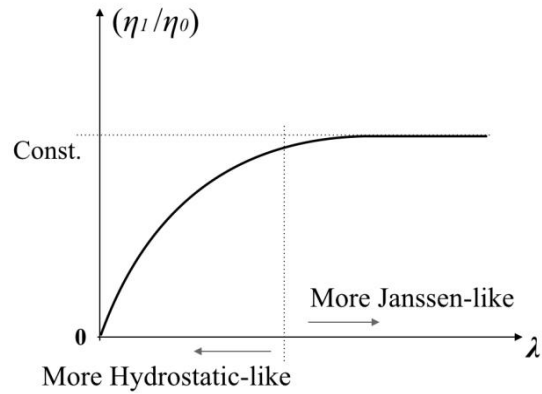

Fig. 5 Schematic of $\eta_1/\eta_0 = g(\lambda)$.

## 3. Experimental observation

The experimental setup is depicted in Fig. 6. The glass beads are dry, non-cohesive, spherical, and are poured into a vertical silo. The density of grain is 2.4 g/cm$^3$. The silo is fixed on a supporting shelf. The bottom of silo is made of a moveable piston. The silo is filled with grains by raining method. After fully relaxation of the initial filling, the piston is allowed to descend at a constant velocity $v$ = 0.02 mm/s on a total distance of 6 cm. This descent of piston could fully mobilize the frictional force between the grains and the wall, and make the average packing fraction to reach a stable value [6, 13]. Five pressure sensors with a uniform interval are inserted into the side-wall so that the head of sensor exactly on the inner surface of silo. After the piston has fully stopped for about 5 minutes, the data are collected. The whole experimental system is at room temperature $25 \pm 1°C$.

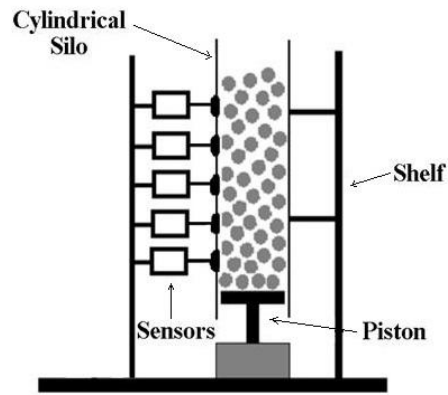

Fig. 6 Schematic of experimental setup

The experiments are performed with different diameters of silos and beads. The results are shown in figs. 7 and 8, respectively, including experimental data and theoretical curves of our model, Rankine's theory, hydrostatics theory and Jansen model. Obviously, Rankine's theory and Janssen's model are all invalid to express experimental results, while our model is in good agreement with it.

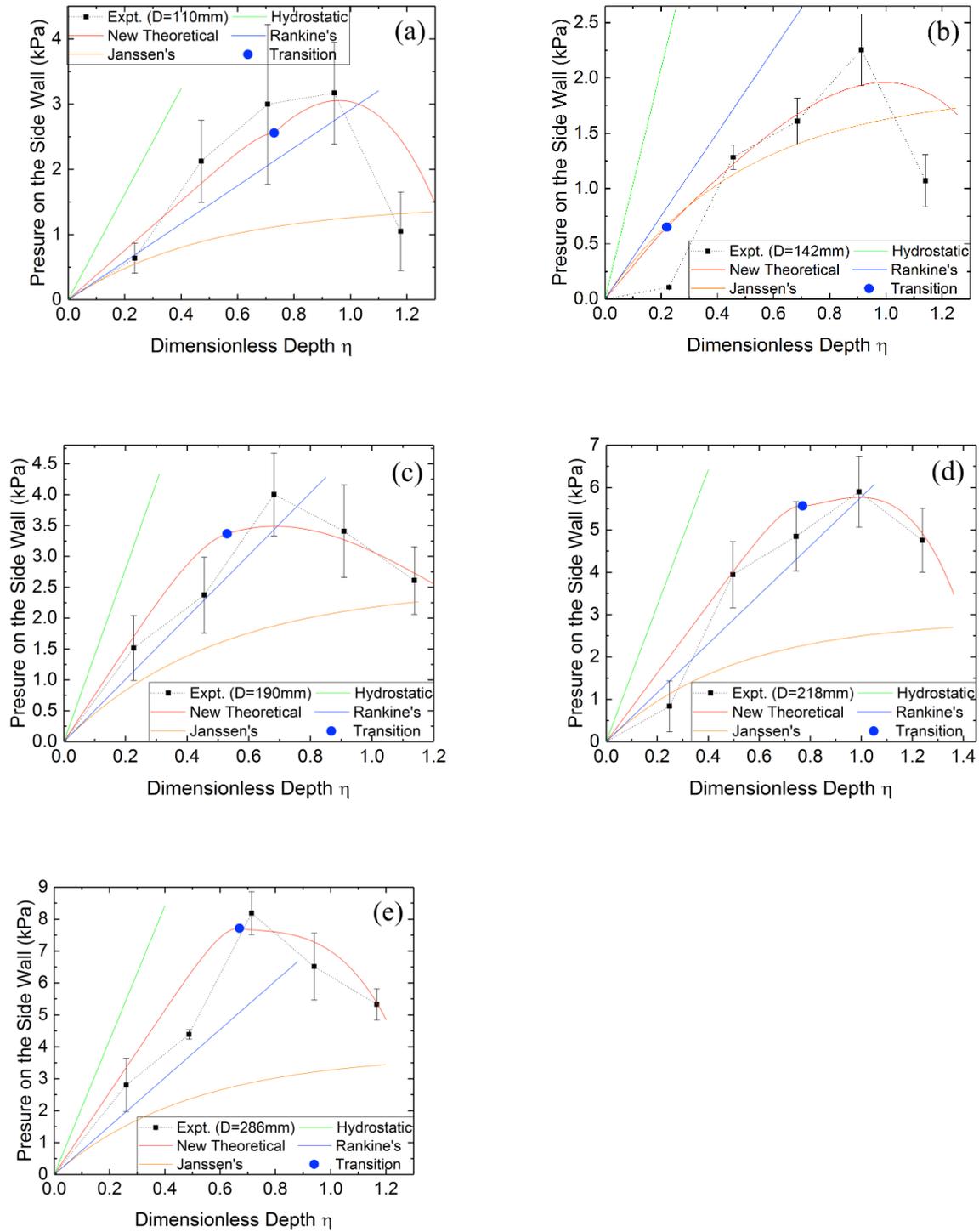

Fig. 7 Experimental data of glass silo vs. glass beads, and theoretical curves of our model, Rankine's theory, hydrostatic theory, and Janssen's model. (a) $D$=110$mm$, $a$=6$mm$, $H$=440$mm$; (b) $D$=142$mm$, $a$=1.5$mm$, $H$=550$mm$; (c) $D$=190$mm$, $a$=2$mm$, $H$=730$mm$; (d) $D$=218$mm$, $a$=2.5$mm$, $H$=917$mm$; (e) $D$=286$mm$, $a$=3$mm$, $H$=1060$mm$.

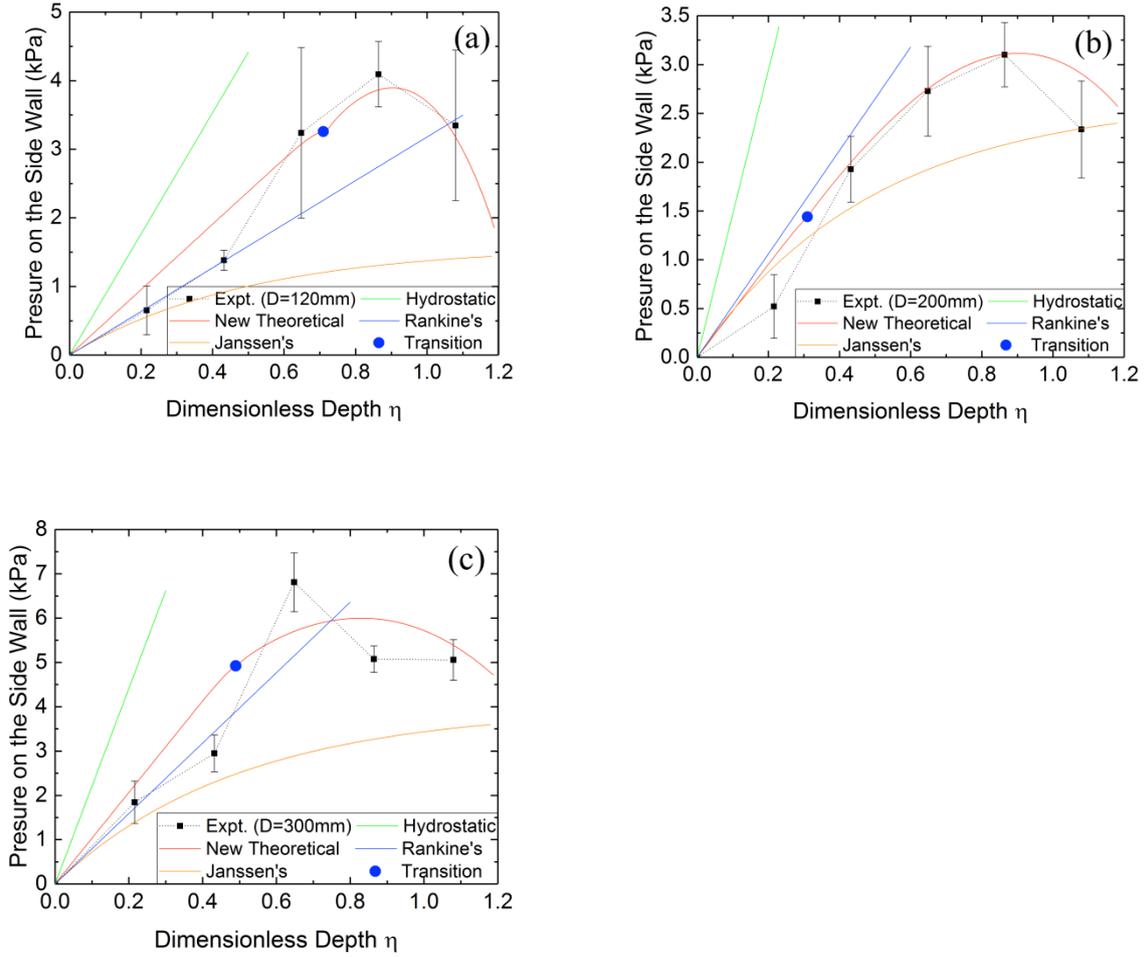

Fig. 8 Experimental data of graphite silo vs. glass beads, and theoretical curves of our model, Rankine's theory, hydrostatic theory, and Janssen's model. (a) *D*=120*mm*, *a*=6*mm*, *H*=440*mm*; (b) *D*=200*mm*, *a*=2*mm*, *H*=550*mm*; (c) *D*=300*mm*, *a*=3*mm*, *H*=730*mm*;

Furthermore, Fig. 7 reveals that with the increase of diameter D of glass silo, the curve of side wall pressure distribution tends to be hydrostatic in upper region, while in bottom region always tend to be Janssen-like. In Fig. 8, silos are made of graphite, and the influence of a different friction coefficient cannot be ignored (i.e. coefficients *q*, *b*, *m* are different from previous experiments in glass silos). However, the results show that the wall material has no essential effect on the general trend of experimental curve.

Now, we re-investigate the transition between hydrostatic-like state to Janssen-like state, i.e., the effect of $\lambda$ on the ratio $\eta_1/\eta_0$ in eq. 21. It is found that theoretical expression is in accordance with experimental results as shown in Fig. 9. In addition, factor *B* is related to the coupling effect varied for different particles and geometric settings, and can be determined experimentally.

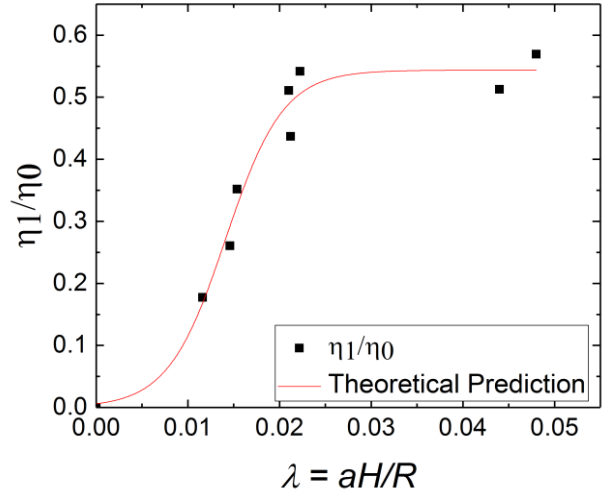

Fig. 9 Fitting result of $\eta_1/\eta_0$ with eq. (21), and parameters used are A=318.427, B=0.54365, $\lambda_0$=0.1412.

## 4. CONCLUSION

A new theoretical model for the side-wall pressure in silos is developed through modifying Janssen's hypothesis of stress ratio. Owing to the hydrostatic-like state in upper region and the geometric restriction in bottom region, the stress states of granular particles located at different positions are non-uniformed. So, it is reasonable to assume stress ratio and friction coefficient changing with depth position when analyzing the distribution of side-wall pressure. Besides, a simple polynomial with five changeable parameters is used to analog this behavior and is shown to be effective for explaining the existence of pressure peak. Furthermore, through the model presented and experiments performed the transition of stress state between hydrostatic-like and Janssen-like is systematically investigated. Finally an empirical expression is proposed to describe this transition phenomenon. Our results are helpful not only for the storage of granular materials in industry, but also for the understanding of mechanical properties of granular materials.

# ACKNOWLEDGMENTS

This work was supported by the Chinese National Science Foundation, Project Nos. 10975014, 11727801 and 11772053.

# REFERENCES

[1] A. Dogangun, Z. Karaca, A. Durmus, and H. Sezen, J. Perf.Construct. Facil. 23, 65 (2009).

[2] B. Dutta, Global J. Res. Anal. 2, 41 (2013).

[3] Huber-Burnand, Ann. Phys. 92, 316 (1829).

[4] W. Rankine, Phil. Transc. Roy. Soc. 47, 9-27 (1857).

[5] H. Janssen, Z. Ver. Deut. Ing. 39, 1045 (1895).

[6] L. Vanel, P. Claudin, J. P. Bouchaud, M. E. Cates, E. Clement, and J. P. Wittmer, Phys. Rev. Lett. 84, 1439 (2000).

[7] Ge B. L. , Shi Q. F. , Ram C. , et al, Chinese Physics Letters, 2013, 30(4):048101.

[8] Wittmer, J. P., et al, Nature 382.6589(1996):336-338.

[9] Jamieson, J. A., 1904, Canadian Society of Civil Engineers, Vol. 17, pp. 554-607

[10] Zhu H. P., Yu A. B., Granular Matter, 2005, 7(2-3):97-107.

[11] Jenike A. W., Johanson J. R., Carson J. W., Journal of Engineering for Industry, 1973, 95(1):1.

[12] Rodolfo Blanco-Rodríguez, and Gabriel Pérez-Ángel, Phys. Rev. E 97.1(2018).

[13] L. Vanel and E. Clément, Eur. Phys. J. B 11, 525 (1999)

[14] Lvin, Powder Technology 4, 280 (1971).

# APPENDIX Symbols used in this paper

| Symbol | Implication | Symbol | Implication |
| --- | --- | --- | --- |
| $p_v$ | Vertical stress | $r$ | Inner radius of ring |
| $p_h$ | Side-wall pressure | $\psi$ | $\psi = kp_v/\gamma R$ |
| $y$ | Depth position | $\Psi$ | $\Psi = \rho\psi$ |
| $\gamma$ | Specific weigh | $\eta$ | Dimensionless depth position $\eta = ky/R$ |
| $R$ | Inner radius of silo | $\rho$ | Dimensionless inner radius of ring $\rho = r/R$ |
| $k$ | $fJ_0$ | $\phi$ | Internal angle friction of the glass beads |
| $K$ | Transferring ratio $K=J(\eta)\mu(\eta)$ | $\eta_0$ | $\eta_0 = kH/R$ |
| $f$ | Friction coefficient of granular beads | $\eta_1$ | Length of the bottom region |
| $\mu$ | Average friction coefficients in the granular column | $m, s, q, b$ | Parameters of our model |
| $J_0$ | Janssen's stress ratio | $\lambda$ | $\lambda = \dfrac{aH}{R}$ |
| $J$ | Stress ratio used in our model | $A, B, \lambda_0$ | Parameters of the expression of the stress state transition |